\documentclass[twocolumn,twoside,preprintnumbers,amsmath,amssymb,showkeys]{revtex4}

\usepackage{epsfig}

\usepackage{graphicx}

\usepackage{fancyhdr}
\usepackage{pslatex}

\begin{document}
\title{Femtoscopic correlations in multiparticle
production and beta-decay}
\author{R. Lednick\'y$^{1,2}$}

\affiliation{$^1$ Joint Institute for Nuclear
Research, Dubna, Moscow Region, 141980, Russia\\
$^2$ Institute of Physics ASCR,
Na Slovance 2, 18221 Prague 8, Czech Republic}

\begin{abstract}
The basics of formalism of femtoscopic and spectroscopic correlations are given,
the orthogonal character of these correlations is stressed.
The similarity and difference of femtoscopic correlations
in multiparticle production and beta-decay is discussed.

\end{abstract}

\vskip -1.35cm

\maketitle

\thispagestyle{fancy}

\setcounter{page}{1}

\bigskip

\section{INTRODUCTION}

The momentum correlations of two or more particles at small
relative momenta in their center-of-mass (c.m.) system
are widely used to study space-time characteristics of the
production processes on a level of fm $=10^{-15}$ m,
so serving as a correlation femtoscopy tool.
Particularly, for non-interacting identical particles,
like photons or,
to some extent, pions, these correlations result from the
interference of the production amplitudes due to the
symmetrization requirement of quantum statistics
(QS) \cite{GGLP60,gkp71,KP72,kop74,pod89}.

The momentum QS correlations were first
observed as an enhanced production of the
pairs of identical pions with small opening angles
(GGLP effect \cite{GGLP60}). Later on,
Kopylov and Podgoretsky
settled the basics of correlation femtoscopy
in more than 20 papers (see a review \cite{pod89})
and developed it as a practical tool;
particularly,
they suggested to study the interference effect
in terms of the correlation function,
proposed the mixing techniques to construct the uncorrelated
reference sample
and clarified the role
of the space--time characteristics
of particle production in various physical situations.

There exists \cite{gkp71,KP72,kop74,pod89,shu73,coc74,KP75}
an analogy of the momentum QS
correlations of photons with the space--time correlations
of the intensities of classical electromagnetic fields used in
astronomy to measure the angular radii of stellar objects
based on the superposition principle (HBT effect) \cite{hbt}.
This analogy is sometimes misunderstood and the momentum
correlations are mixed up with the space-time (HBT)
correlations in spite that their orthogonal character
and thus the absence of the former in astronomy measurements
due to extremely large space-time extent of stellar objects
(and vice versa) was already pointed out in early
paper \cite{KP75} (see also \cite{lyu98}).

Note that though the space-time (HBT) correlations
are absent in subatomic measurements, they
can still be used in laboratory
as an intensity-correlation spectroscopy
tool allowing one to measure the spectral line
shape and width (see \cite{glw66,pkd67} and references therein).
In fact, the QS space-time correlations give the
information about shape and width of the three-momentum
distribution (including the angular one) of the quanta
coming from a distant source and can be considered
as a spectroscopy tool in this more general sense.

The momentum correlations of particles emitted at nuclear
distances are also influenced by the effect of
final state interaction (FSI) \cite{koo,gkw79,ll1,BS86,bgj90}.
Thus the effect of the Coulomb interaction dominates the correlations
of charged particles at the relative momenta $2k^*$ in the
two-particle rest frame smaller than
the inverse Bohr radius
$|a|^{-1}$ of the two-particle system,
respectively suppressing or
enhancing the production of particles with like or unlike charges.

Though the FSI effect complicates the correlation analysis,
it is an important source of information allowing for the
coalescence femtoscopy (see, {e.g.}, \cite{sy81,lyu88,mro93,sh99}),
the correlation femtoscopy with unlike particles \cite{ll1,BS86,bgj90}
including the access to the relative space--time
asymmetries in particle production
\cite
{LLEN95,ali95,vol97,sof97,lpx,led03,mis98,led02,blu02,ada03a,kis04,led04}
and a study of strong interactions between specific particles
\cite{led02,kis04,led04}.

In fact, the femtoscopic Coulomb correlations were observed
more than 70 years ago when a sensitivity of the differential
beta-decay rate to the nucleus charge and radius
has been established \cite{fer34}.

The paper is organized now as follows.  In sections II-IV we
give the basics of formalism of
femtoscopic and spectroscopic correlations.
The similarity and difference of femtoscopic correlations
in multiparticle production and beta-decay is discussed
in section V. We conclude in section VI.
For recent results from femtoscopy of multiparticle
production processes,
one can inspect reviews \cite{led04,lis05,cso05,led06}.

\section{FEMTOSCOPIC QS CORRELATIONS}
\subsection{Formalism}

The correlation function
${\mathcal R}(p_{1},p_{2})$ of
two particles with four-momenta $p_1$ and $p_2$
is usually defined as the
ratio of the measured momentum distribution
of the two particles to the
reference one obtained by mixing particles from different
events of a given class,
normalized to unity at sufficiently
large relative momenta.

For identical pions or kaons, the effect of the strong FSI is usually
small and the effect of the Coulomb FSI can be in first approximation
simply corrected for (see \cite{slape} and references therein).
The corrected correlation function is then determined by the
QS symmetrization only.
The space-time information contained in the momentum QS
correlations can be extracted based on the
Kopylov-Podgoretsky (KP) formalism of independent one-particle
classical emitters that are characterized by their
production (excitation)
four-coordinates, four-velocities, lifetimes and decay amplitudes.
The KP formalism is valid on the following assumptions:

(i) The multiplicity of produced particles is assumed sufficiently
large to neglect the influence of conservation laws.

(ii) An independent or incoherent particle emission is assumed,
i.e. the coherent contribution of multiparticle emitters is neglected
(this contribution can be eventually taken into account with the
help of the suppression parameter $\lambda$).

(iii) The mean freeze-out phase space density
is assumed sufficiently
small so that the correlation function of two particles
emitted with a small relative momentum $Q=2 k^*$ in their
c.m. system is mainly determined by their mutual correlation.

(iv) The momentum dependence of the one-particle emission
probabilities is assumed
inessential when varying the particle four-momenta
$p_{1}$ and $p_{2}$ by the
amount characteristic for the correlation effect.
This so-called {\it smoothness} assumption requires the components of the
mean space-time distance between particle emitters
much larger than those of the space-time extent
of the emitters.

The probability amplitude to observe a particle with the four-coordinate $x$
from an emitter A decaying at the four-coordinate $x_{\rm A}$
can depend on $x$ through the relative four-coordinate $x-x_{\rm A}$
only and so can be written in the form:
\begin{equation}
\langle x|\psi_{\rm A}\rangle = (2\pi)^{-4}
\int d^4\kappa~u_{\rm A}(\kappa)\exp[i\kappa(x-x_{\rm A})].
\label{Px}
\end{equation}
We assume here that after the production the emitter moves
along a classical trajectory and decays
according to the exponential decay law.
Such a classical treatment of the decay is
often applied also to resonances (see, e.g., \cite{Led92,ll96,cp01}).
It is justified for
a heavy emitter with the energy spectrum of the decay particle
substantially wider than the emitter width.
One can avoid the classical treatment of the emitter decay by
considering $x_{\rm A}$ in Eq. (\ref{Px})
as the emitter production (excitation) four-coordinate and adding,
in the case of a negligible emitter space-time extent, the
theta-function $\theta(t-t_{\rm A})$.

The probability amplitude to observe the particle with
the four-momentum $p$ is
\begin{equation}
\langle p|\psi_{\rm A}\rangle =
\int d^4 x~\langle p|x\rangle \langle x|\psi_{\rm A}\rangle
= u_{\rm A}(p)\exp(-i p x_{\rm A}),
\label{Pp}
\end{equation}
where $\langle p|x\rangle=\exp(-i p x)$.
The probability amplitude to observe identical spin-0 bosons with
four-momenta $p_1$ and $p_2$ emitted by emitters A and B
should be symmetrized in accordance with the requirement of QS:
\begin{equation}
T_{{\rm A}{\rm B}}^{\rm sym}(p_1,p_2)=
[\langle p_1|\psi_{\rm A}\rangle \langle p_2|\psi_{\rm B}\rangle +
\langle p_2|\psi_{\rm A}\rangle \langle p_1|\psi_{\rm B}\rangle
]/\sqrt{2}.
\label{Pp1p2}
\end{equation}
The corresponding QS correlation function is given by the
properly normalized square of this amplitude
averaged over all
characteristics of the emitters:
\begin{eqnarray}
&&{\mathcal R}(p_{1},p_{2}) = 1
\nonumber\\
&&\hskip0.6cm +\frac{\Re \sum_{\rm AB} u_{\rm A}(p_1)u_{\rm B}(p_2)
u_{\rm A}^*(p_2)u_{\rm B}^*(p_1)
\exp(-i q \Delta x)}
{\sum_{\rm AB}|u_{\rm A}(p_1)u_{\rm B}(p_2)|^2}
\nonumber\\
&&\hskip1.5cm \doteq 1+\langle \cos(q\Delta x) \rangle ,
\label{CF1}
\end{eqnarray}
where $q=p_1-p_2$, $\Delta x= x_{\rm A}-x_{\rm B}$
and the last equality follows from the
{\it smoothness} assumption (iv).

It should be noted that the last equality in Eq. (\ref{CF1})
is usually used to calculate correlation
functions within classical transport models identifying the emitter
four-coordinates as those of the decay four-coordinates
of the primary emitters including
resonances or those of
the last collisions of the emitted particles.
Concerning the accuracy of the classical approach to the
emitter decay, we note that, for example
in the case of a $\rho$-meson and a pion emitted from a small
space-time region, it is a valid approximation at $Q<0.1$ GeV/$c$
but overestimates the tail of the corresponding two-pion
correlation function by about 15 percent
(see figure 2 in \cite{Led92}).
This overestimation is however not important when the space-time
separation of the emitters is larger than 2 fm
(as in heavy ion collisions) and so the interference effect
rapidly vanishes at $Q>0.1$ GeV/$c$.

It is instructive to introduce the emission function
$G(\bar{x},p)$
(similar to Wigner function)
as a partial Fourier transform of the
space-time density matrix:
\begin{equation}
G(\bar{x},p)=
\int d^4\epsilon~\exp(-i p \epsilon)\sum_{\rm A}
\langle \bar{x}+\frac12\epsilon|\psi_{\rm A}\rangle
\langle \psi_{\rm A}|\bar{x}-\frac12\epsilon\rangle .
\label{G}
\end{equation}
Since the single-particle emission probability is given
by the integral
\begin{equation}
\int d^4\bar{x}~G(\bar{x},p)=
\sum_{\rm A}|\langle p|\psi_{\rm A}\rangle|^2 ,
\label{P1}
\end{equation}
the emission function, though not positively defined,
can be usually interpreted as a probability density
to find a particle with four-momentum $p$ and an average
four-coordinate $\bar{x}=\frac12(x+x')$.
For the QS correlation function of two identical
spin-0 bosons one has:
\begin{eqnarray}
&&{\mathcal R}(p_{1},p_{2}) = 1
+\frac{\int d^4 \bar{x}_1 d^4 \bar{x}_2~
G(\bar{x}_1,p)G(\bar{x}_2,p)\cos(q\Delta \bar{x})}
{\int d^4 \bar{x}_1 d^4 \bar{x}_2~
G(\bar{x}_1,p_1)G(\bar{x}_2,p_2)}
\nonumber\\
&&\hskip1.5cm \doteq 1+\langle \cos(q\Delta \bar{x}) \rangle ,
\label{CF2}
\end{eqnarray}
where $\Delta \bar{x}=\bar{x}_1-\bar{x}_2$ and
$p=\frac12(p_1+p_2)\equiv \frac12 P$.
Similar to Eq. (\ref{CF1}),
the last equality follows from the
{\it smoothness} assumption allowing one
to identify the average four-coordinate $\bar{x}$
of the emitted particle
with the decay four-coordinate of its emitter,
i.e. neglect the space-time extent of the emitter.

Generally, for two identical bosons or fermions with
the total spin $S$, the "+" sign in Eq. (\ref{Pp1p2})
should be substituted by $(-1)^S$ or, equivalently, the
two-particle amplitude should be symmetrized (antisymmetrized)
only for identical bosons (fermions) emitted with the
same spin projections.
As a result, in the case of initially unpolarized spin-$j$
particles, the "+" sign in Eq. (\ref{CF1}) or (\ref{CF2})
for the QS correlation function
should be substituted by $(-1)^{2j}/n_j$,
where $n_j$ is the number of possible spin projections:
$n_j=2j+1$ or, for massless particles, $n_j=2$.

A special comment requires the correlation function of
massless particles with the spin $j > 1/2$ when the
helicities in-between the extreme values $\pm j$ are
forbidden. Thus, for photons,
due to the transversality of the electromagnetic
field (forbidden zero helicity),
the spin factor multiplying the correlation term
$\langle \cos(q\Delta \bar{x}) \rangle$ actually depends on
the angle $\theta$ between the photon three-momenta
\cite{neu86,lp95}.
Particularly, in the model of randomly oriented one-photon
dipole or quadrupole emitters, it equals $\frac14(1+\cos^2\theta)$
or $\frac14(1-3\cos^2\theta+4\cos^4\theta)$, respectively
\cite{lp95}. However, in the case of practical interest,
when the photon wavelength is essentially smaller than the
size of the photon emission region, the angle $\theta$
between photons contributing to the interference effect
is very small and the spin factor reduces to the value
$1/n_j=\frac12$ irrespective of the multipole order
of the emitter.

A comment requires also the correlation function of
neutral kaons. Two neutral kaons with four-momenta $p_1$
and $p_2$ are originally produced
as pairs $K^0(p_1)K^0(p_2)$, $\bar{K}^0(p_1)\bar{K}^0(p_2)$,
$K^0(p_1)\bar{K}^0(p_2)$ and $\bar{K}^0(p_1)K^0(p_2)$.
The correlation pattern now depends on the way the neutral
kaons are detected. In principle, one can detect $K^0$ and
$\bar{K}^0$, e.g., by charge exchange reactions $K^0\to K^+$ and
$\bar{K}^0\to K^-$. In this case, only the first two
production channels would give the interference effect
similar to that for identical pions.  Usually, however,
neutral kaons are detected through their two-pion and
three-pion decays as so-called short-lived ($K_S^0$) and
long-lived ($K_L^0$) states, respectively. Neglecting the
small effect of $CP$-violation, these states correspond to the
eigen states with $CP=\pm 1$:
\begin{equation}
|K_S^0\rangle=(|K^0\rangle+|\bar{K}^0\rangle)/\sqrt{2},~~~
|K_L^0\rangle=(|K^0\rangle-|\bar{K}^0\rangle)/\sqrt{2}.
\end{equation}
Therefore, all four production channels of the pairs
of $K^0$- and $\bar{K}^0$-mesons contribute to the production
of the pairs $K_S^0(p_1)K_S^0(p_2)$, $K_L^0(p_1)K_L^0(p_2)$,
$K_S^0(p_1)K_L^0(p_2)$ and $K_L^0(p_1)K_S^0(p_2)$.
It is clear that the channels
$K^0(p_1)K^0(p_2)$ and $\bar{K}^0(p_1)\bar{K}^0(p_2)$ yield
the constructive interference pattern for any combination of the
detected pairs of $K_S^0$- and $K_L^0$-mesons.
It is interesting that also the channels
$K^0(p_1)\bar{K}^0(p_2)$ and $\bar{K}^0(p_1)K^0(p_2)$ now
yield the interference effect which can be both constructive
and destructive. Indeed, since there are two indistinguishable
amplitudes contributing to final state, the symmetrized
amplitude has the form \cite{lp79}:
\begin{eqnarray}
&&T_{{\rm A}{\rm B}}^{\rm sym}(K_i^0(p_1),K_j^0(p_2))
\nonumber\\
&&\hskip.3cm =
~~\langle K_i^0|K^0\rangle\langle K_j^0|\bar{K}^0\rangle
\langle p_1|\psi_{\rm A}(K^0)\rangle \langle p_2|\psi_{\rm B}(\bar{K}^0)\rangle
\nonumber\\
&&\hskip.5cm +
\langle K_j^0|K^0\rangle\langle K_i^0|\bar{K}^0\rangle
\langle p_2|\psi_{\rm A}(K^0)\rangle \langle p_1|\psi_{\rm B}(\bar{K}^0)\rangle,
\label{PKKp1p2}
\end{eqnarray}
where the sign "+" corresponds to Bose statistics
of neutral kaons. One may see that the corresponding interference
pattern is constructive for $K_S^0K_S^0$- and $K_L^0K_L^0$-pairs
while it is destructive for $K_S^0K_L^0$-pairs.

\subsection{Simple parameterizations}

A characteristic feature of the QS correlation function
of two identical bosons (fermions) is
the presence of the interference maximum (minimum) at small
components of the relative four-momentum $q$
with the width reflecting the inverse space-time extent
of the effective production region.
For example, assuming that for a
fraction $\lambda$ of pion pairs,
the pions are emitted independently according to
one--particle amplitudes of a Gaussian form characterized by the
space--time dispersions $r_0^2$ and $\tau_0^2$
while, for the remaining fraction $(1-\lambda)$
related to very long--lived emitters ($\eta$, $\eta'$, $K^0_s$,
$\Lambda$, \dots), the relative distances $r^*$ between the emitters
in the pair c.m. system are extremely large,
one has
\begin{eqnarray}
&&{\cal R}(p_{1},p_{2})=
1+\lambda\exp\left(-r_0^2{\bf q}^2-\tau_0^2q_0^2\right)
\nonumber\\
&&\hskip1.5cm =
1+\lambda\exp\left(-r_0^2{\bf q}_T^2-(r_0^2+v^2\tau_0^2)q_L^2\right),
\label{5}
\end{eqnarray}
where $q_T$ and $q_L$ are the transverse and longitudinal
components of the three--momentum difference ${\bf q}$
with respect to the direction
of the pair velocity ${\bf v}={\bf P}/P_0$.
One may see that, due to the on-shell constraint \cite{KP72}
$q_0 = {\bf v}{\bf q} \equiv v q_L$
(following from the equality $qP=0$), strongly correlating
the energy difference $q_0$ with the longitudinal momentum
difference $q_L$,
the correlation function at $v\tau_0 > r_0$ substantially depends
on the direction of the vector  ${\bf q}$
even in the case of a
spherically symmetric spatial form of the production region.

Note that the on-shell constraint makes the $q$-dependence
of the correlation function essentially three--dimensional
and thus makes impossible the unique Fourier reconstruction
of the space--time characteristics of the emission process.
Particularly, in pair c.m. system,
$q = \{0,2{\bf k}^*\}$, $\Delta x = \{t^*,{\bf r}^*\}$) and
the scalar product $q\Delta x=-2{\bf k}^*{\bf r}^*$
is independent of the time difference $t^*$.
However, within realistic models,
the directional and velocity
dependence of the correlation function
can be used to determine both
the duration of the emission and the form
of the emission region \cite{KP72}, as well as - to reveal
the details of the
production dynamics (such as collective flows;
see, {\it e.g.}, \cite{PRA84,MAK87} and the reviews
\cite{WH99,cso02}).
For this, the correlation functions can be analyzed
in terms of the {out} (x), {side} (y) and {longitudinal} (z)
components of the relative momentum vector
${\bf q}=\{q_x,q_y,q_z\}$ \cite{pod83,osl-sys};
the {out} and {side} denote the transverse,
with respect to the reaction axis, components
of the vector ${\bf q}$, the {out} direction is
parallel to the transverse component of the pair three--momentum.
The corresponding correlation widths are
usually parameterized in terms
of the Gaussian correlation radii $R_i$,
\begin{equation}
{\cal R}(p_{1},p_{2})=
1+\lambda\exp(-R_x^2q_x^2-R_y^2q_y^2-R_z^2q_z^2
-R_{xz}^2q_xq_z)
\label{osl}
\end{equation}
and their dependence on pair rapidity and transverse momentum
is studied.
The form of Eq.~(\ref{osl}) assumes azimuthal symmetry of the
production process \cite{WH99,pod83}. Generally, {\it e.g.},
in case of the
correlation analysis with respect to the reaction plane, all three
cross terms $q_iq_j$ contribute \cite{wie_fi98}.

It is well known that particle correlations at high energies
usually measure only a small part of the space-time emission volume,
being only slightly sensitive to its increase related to the fast
longitudinal motion of particle emitters. In fact,
due to limited emitter decay momenta
of few hundred MeV/c, the correlated particles with nearby velocities
are emitted by almost comoving emitters and so - at nearby space--time
points.
The dynamical examples are resonances,
colour strings or hydrodynamic expansion.
To substantially eliminate the effect of the longitudinal motion,
the correlations can be
analyzed in terms of the invariant variable
$Q = (-{q}^2)^{1/2} = 2k^*$ and
the components of the  momentum difference in pair c.m. system
(${\bf q}^*\equiv {\bf Q}= 2{\bf k}^*$) or in
the longitudinally comoving system (LCMS) \cite{cso91}.
In LCMS each pair is emitted transverse to the reaction axis
so that the relative momentum
${\bf q}$ coincides with ${\bf q}^*$
except for the component
${q}_x=\gamma_{t}q_x^*$,
where $\gamma_{t}$ is the LCMS Lorentz factor of the pair.

Particularly, in the case of one--dimensional boost invariant
expansion, the longitudinal correlation radius in the LCMS reads
\cite{MAK87} $R_z \approx (T/m_t)^{1/2}\tau$, where $T$ is the
freeze-out temperature, $\tau$ is the proper freeze-out time and
$m_t$ is the transverse particle mass.
In this model, the side radius measures the
transverse radius of the system while,
similar to Eq.~(\ref{5}), the square of the out radius
gets an additional contribution $(p_t/m_t)^2\Delta\tau^2$
due to the finite emission duration $\Delta\tau$.
The additional transverse expansion leads to a slight
modification of the $p_t$--dependence of the longitudinal radius and -
to a noticeable decrease of the side radius and the spatial
part of the out radius with $p_t$.
Since the freeze-out temperature and the transverse flow determine
also the shapes of the $m_t$-spectra,
the simultaneous analysis of correlations and single particle
spectra for various particle species allows one to disentangle
all the freeze-out characteristics \cite{WH99}.

\section{SPECTROSCOPIC QS CORRELATIONS}

To help in understanding the analogy and difference
of the QS space-time (spectroscopic) and momentum
(femtoscopic) correlations,
we briefly present here the formalism of QS correlation
spectroscopy within the KP model of independent single-particle
emitters. In the spectroscopic correlation measurements the
particles are supposed to be emitted by a distant object
with large space-time dimensions
and detected by two detectors at space-time points $x_1$ and
$x_2$. It is assumed that the distance between the
detectors is much smaller than the size of the emitting object
and that this size is negligible compared with the distance
between the object and detectors.
Then the four-momentum of a photon emitted by the emitter A
and detected by any of the two detectors can be written
as $p_{\rm A}=\omega_{\rm A}\{1,\hat{p}_{\rm A}\}$, where
$\hat{p}_{\rm A}$ is the unit vector in the direction
from the emitter A to the detectors.
The four-dimensional integral in the single-photon
probability amplitude in Eq. (\ref{Px}) then reduces to
the one-dimensional one:
\begin{equation}
\langle x|\psi_{\rm A}\rangle = (2\pi)^{-1}
\int d\omega_{\rm A}~u_{\rm A}(\omega_{\rm A})
\exp[i p_{\rm A}(x-x_{\rm A})]\theta(t-t_{\rm A}),
\label{Px1}
\end{equation}
where $u_{\rm A}(\omega_{\rm A})\propto
(\omega_{\rm A}-\omega_{0\rm A}-i\Gamma_{\rm A})^{-1}$, i.e.
the emitter decay is treated quantum-mechanically
and parameterized by the energy $\omega_{0\rm A}$
and width $\Gamma_{\rm A}$ of the emission line.
In accordance with the comment after Eq. (\ref{Px}),
$x_{\rm A}$ now denotes the excitation four-coordinate
of the emitter and the condition $t>t_{\rm A}$ is
introduced by the theta-function.
Since the time $t_{\rm A}$ is distributed in a very wide
interval, the sum $\sum_{\rm A}
\exp[-i(\omega_{\rm A}-\omega_{\rm A}')t_{\rm A}]$
yields the delta-function $\delta(\omega_{\rm A}-\omega_{\rm A}')$
so that the single-photon probability is merely proportional
to the integral of the spectral function:
\begin{equation}
\sum_{\rm A}|\langle x|\psi_{\rm A}\rangle|^2
\propto \sum_{\rm A}\int d\omega_{\rm A}~
|u_{\rm A}(\omega_{\rm A})|^2.
\label{P11}
\end{equation}
The probability amplitude of two photons with the same
and complete polarization
should be symmetrized similar to Eq. (\ref{Pp1p2}):
\begin{equation}
{\cal T}_{{\rm A}{\rm B}}^{\rm sym}(x_1,x_2)=
[\langle x_1|\psi_{\rm A}\rangle \langle x_2|\psi_{\rm B}\rangle +
\langle x_2|\psi_{\rm A}\rangle \langle x_1|\psi_{\rm B}\rangle
]/\sqrt{2}.
\label{Px1x2}
\end{equation}
For photons with polarization $P$, the symmetrized amplitude
(\ref{Px1x2}) describes only the fraction $\frac12 (1+P^2)$
of the photon pairs. As a result,
the correlation function
${\mathcal R}(x_{1},x_{2})$, defined as a number of
two-photon counts normalized to unity at a large space-time
separation of the detected photons, becomes
\begin{eqnarray}
&&{\mathcal R}(x_{1},x_{2}) = 1+\frac{1+P^2}{2}\cdot
\nonumber\\
&&\cdot\frac{\sum_{\rm AB}\int d\omega_{\rm A}d\omega_{\rm B}
|u_{\rm A}(\omega_{\rm A})|^2
|u_{\rm B}(\omega_{\rm B})|^2
\cos(q_{\rm AB}\Delta x)}
{\sum_{\rm AB}\int d\omega_{\rm A}d\omega_{\rm B}
|u_{\rm A}(\omega_{\rm A})|^2
|u_{\rm B}(\omega_{\rm B})|^2}
\nonumber\\
&&\hskip1.5cm = 1+\frac{1+P^2}{2}
\langle\cos(q_{\rm AB}\Delta x) \rangle ,
\label{CF11}
\end{eqnarray}
where $q_{\rm AB}=p_{\rm A}-p_{\rm B}$, $\Delta x=x_1-x_2$.

It should be noted that
the HBT technique is not based on counting separate quanta,
it overcomes this difficult problem
by the measurement of the product of fluctuating parts of the
electric currents from the two detectors
(the low-frequency part is filtered out)
integrated and read in
time intervals of the order of minutes.
To get rid of the uncertainties in the detector
gains the measured quantity is the ratio of the
product mean to root-mean-square deviation.
It can be shown that the correlation coefficient
defined as this ratio normalized to unity at zero
distance between the detectors,
equals to
\begin{equation}
\langle\cos({\bf q}_{\rm AB}\Delta {\bf x}) \rangle\approx
[2J_1(\rho)/\rho]^2.
\label{HBT}
\end{equation}
The approximate equality is valid for a uniformly
radiating disk with the normal directed to the detectors
(or, for a spherical surface radiating according to Lambert law)
emitting light of a small band width;
the argument of Bessel function $J_1$ is
$\rho=\bar{\omega}\theta d$, where $\bar{\omega}$ is
the mean angular frequency (mean energy of the detected photons),
$d$ is the distance between the detectors perpendicular to the
direction to the distant emitting object and $\theta$ is the
object angular radius. Measuring the correlation coefficient
as a function of the distance $d$, one thus determines
the transverse spread $\bar{\omega}\theta$ of the wave vectors
of the detected light (the spread of the transverse photon momenta).
Obviously, the HBT correlation effect is insensitive to the
actual space-time extent of the source. At most, one can
determine the source angular radius $\theta$ performing the
additional spectral measurements of
the mean angular frequency $\bar{\omega}$.

It is interesting to note that Eq. (\ref{HBT}) follows also from
the superposition principle applied to classical electromagnetic fields
and so the HBT intensity correlation effect would survive even in the
case of vanishing
Planck constant when the QS correlations become unobservable.

Comparing the QS space-time correlation function in Eq. (\ref{CF11})
with the QS momentum correlation function in Eq. (\ref{CF1}),
one may see a peculiar symmetry: one is transformed to the other
by the interchange of the emitters and detectors \cite{KP75}.
Thus the space-time correlations yield the momentum picture
of the source while the momentum correlations provide the
information about the source space-time characteristics.

Some historical remarks are appropriate here. The analogy of
QS momentum correlations with the HBT space-time correlations
was first mentioned in paper \cite{gkp71}. However, not stressing
the differences, this paper triggered a number of misleading
statements, such as \cite{shu73}:
"{\it The interest to correlations of identical
quanta is due to the fact that their magnitude is connected with
the space and time structure of the source of quanta. This idea
originates from radio astronomy and is the basis of Hanbury-Brown
and Twiss method of the measurement of star radii}".
To clarify the situation, Kopylov and Podgoretsky wrote a special
paper \cite{KP75} in which they clearly stressed the difference
between the momentum and space-time QS correlation measurements.
Particularly, they pointed out: "{\it when any of the time parameters
characterizing radiating system becomes very large, the possibility
to measure the system dimensions practically vanishes since the
interference effect remains only in the unobservable small
region of the energy difference $q_0$. On the other hand, in
astronomy, it appears to be possible to measure angular dimensions
of stars despite their lifetimes can be considered infinitely
large}". Unfortunately, this clarifying paper missed the attention
of a number of experts in the
field of interference correlations. Thus even in reviews on
the subject one can meet the incorrect statement that there is
no principle difference between QS correlations in particle physics
and astronomy, that both are momentum correlations allowing one
to determine spatial dimensions of the emitting object.
An example of such incorrect view of QS
correlations in astronomy is given in figure 1 \cite{gra77}.
Another example of this widespread error is chapter 1.1 in
a review \cite{bgp92}.
\begin{figure}[!htb1]
\begin{center}
\includegraphics*[angle=0, width=8cm]{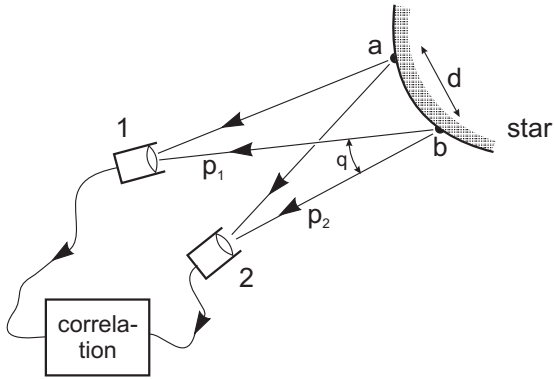}
\end{center}
\caption{\emph{\small An example of the incorrect view
of spectroscopic QS
correlations in astronomy \cite{gra77}.
Neither the extremely small three-momentum difference ${\bf q}$
between the photons from the same emitter
(a negligible part of the photon pairs)
nor the one between the photons from different emitters
can directly be measured in astronomy.
In fact, this figure would be a correct view of femtoscopic
QS correlations if a distant star were substituted by
a nearby object emitting identical particles
at a characteristic space-time distance
of the order of femtometer.}}
\end{figure}

\section{FEMTOSCOPIC FSI CORRELATIONS}

It can be shown \cite{gkw79,ll1,led05}
that the effect of FSI manifests itself
in the fact that the role of functional basis, which the asymptotic
two-particle state is projected on, is transferred from
plane waves $\exp(-ip_1x_1-ip_2x_2)$
to the Bethe-Salpeter amplitudes
$\Psi_{p_1p_2}^{(-)}(x_1,x_2)=\Psi_{p_1p_2}^{(+)*}(x_1,x_2)
=\exp(-iPX)\psi_{\tilde q}^{(+)*}(\Delta x)$, where
$\Delta x\equiv x_1-x_2=\{t,{\bf r}\}$ is
the relative four-coordinate,
$\widetilde{q}=q-P(qP)/P^2$
is the generalized relative four-momentum,
$P=p_1+p_2$ and $qP = m_1{}^2-m_2{}^2$.

To simplify the calculation of the FSI effect,
the Bethe-Salpeter amplitude describing two particles
emitted at space-time points $x_{i}=\{t_{i},{\bf r}_{i}\}$
and detected with four-momenta $p_i$
is usually calculated at equal emission times in the pair
c.m. system;
i.e. the reduced non-symmetrized Bethe-Salpeter amplitude
$\psi_{{\tilde q}}^{(+)}(\Delta x)$
is substituted
in the two-particle c.m. system, where ${\bf P} = 0$,
$\tilde q = \{0,2{\bf k}^*\}$ and $\Delta x = \{t^*,{\bf r}^*\}$),
by a stationary solution
$\psi ^{(+)}_{-{\bf k}^*}({\bf r}^*)$ of the
scattering problem having at large distances
${r}^*$ the asymptotic form of a
superposition of the plane and outgoing spherical waves
(the minus sign of the vector ${\bf k}^{*}$ corresponds to the reverse
in time direction of the emission process).
This {\it equal time} approximation is
valid on conditions \cite{ll1,led05}
$ |t^*|\ll m_{2,1}r^{*2}$ for
${\rm sign}(t^*)=\pm 1$ respectively.
These conditions are usually satisfied
for heavy particles like kaons or
nucleons. But even for pions, the $t^{*}=0$ approximation
merely leads to a slight overestimation (typically $<5\%$) of the
strong FSI effect (see figure 2 and \cite{led05})
and,
it doesn't influence the leading zero--distance
($r^{*}\ll |a|$) effect of the Coulomb FSI.
\begin{figure}[!htb1]
\begin{center}
\includegraphics*[angle=0, width=8cm]{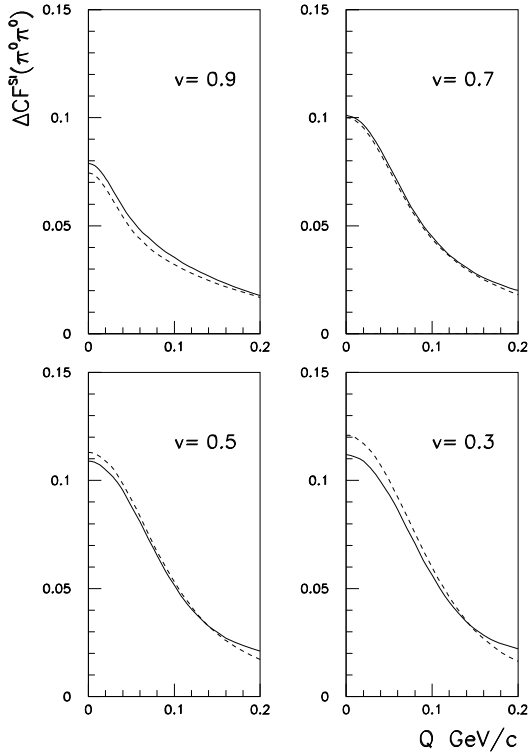}
\end{center}
\caption{\emph{\small The FSI contribution to the $\pi^0\pi^0$ correlation
function calculated for the pair velocity
$v=0.3, 0.5, 0.7, 0.9~c$ in a model of independent
single--particle emitters distributed
according to a Gaussian law with the spatial and time width
parameters $r_0=2$ {\rm fm} and $\tau_0=2$ {\rm fm}/$c$. The exact results
(solid curves) are compared with those obtained in the
equal--time approximation (dash curves).}}
\end{figure}

Note that in the case of small $k^*$, we are interested in,
the short-range interaction is dominated by central forces and
s-waves so that, neglecting a weak spin dependence of the
Coulomb interaction, the spin dependence of the
two-particle amplitude enters only through the total spin $S$.

On the assumptions (i)-(iv),
the two-particle correlation function then reduces to
the square of the two-particle wave function
averaged over the distance ${\bf r}^{*}$ of the emitters
in the two-particle c.m. system
and the total spin $S$ of the pair:
\begin{equation}
{\cal R}(p_{1},p_{2})\doteq
\langle |\psi_{-{\bf k}^{*}}^{(+)}({\bf r}^{*})|^{2}
\rangle .
\label{1}
\end{equation}
For identical particles,
the amplitude in Eq.~(\ref{1})
enters in a symmetrized form:
\begin{equation}
\label{sym}
\psi_{-{\bf k}^{*}}^{(+)}({\bf r}^{*}) \rightarrow
[\psi_{-{\bf k}^{*}}^{(+)}({\bf r}^{*})+(-1)^{S}
\psi_{{\bf k}^{*}}^{(+)}({\bf r}^{*})]/\sqrt{2}.
\end{equation}

The {\it two-particle approximation} in (i) and FSI factorization
in the Bethe-Salpeter amplitudes of the elastic transitions
$1+2\to 1+2$ implies a long FSI time as compared with the
characteristic production time, i.e. the channel momentum $k^*$
much less than typical production momentum transfer of hundreds
MeV/$c$. In fact, the long-time FSI can be also factorized in the
inelastic transitions, $1+2\to 3+4$, characterized by a slow
relative motion in both entrance and exit channels
\cite{lll97,led05}. The necessary
condition is an approximate equality of the sums of particle
masses in the channels $1+2$ and $3+4$.
In the presence of such transitions
the two-channel scattering
problem has to be solved and both elastic and inelastic
transition amplitudes should be taken into account in
the averaging in Eq.~(\ref{1}).
In practice, the particles $1,3$ and $2,4$ are members of the
same isomultiplets (as, e.g., in the transition
$\pi^- p \to \pi^0 n$ or $K^+K^- \to K^0\bar{K^0}$)
so that one can assume the same weights and same
${\bf r}^{*}$-distributions for the channels $1+2$ and $3+4$.

In heavy ion collisions, the effective radius $r_0$ of the
emission region
can be considered much larger than the range of the
strong interaction potential.
The short range FSI contribution to the correlation function
is then independent of the actual
potential form \cite{ll1,gkll86}. At small $Q=2k^*$,
it is determined by the s-wave
scattering amplitudes $f^S(k^*)$ at a given total spin $S$
scaled by the radius $r_0$ \cite{ll1}.
For two-nucleon systems, the scattering lengths $f^S(0)$ are
large (up to $\sim 20$ fm) and
this contribution often dominates over the effect of QS.
For two-meson or meson-baryon systems, the scattering amplitudes
are usually quite small ($< 0.2$ fm) and the short range FSI
contribution (including the contribution of the
coupled channel which is quadratic in the amplitude of the
corresponding inelastic transition)
can be often neglected.
This contribution cannot be however neglected
for the $K\bar{K}$-system due to rather large
s-wave $K\bar{K}$ scattering length
dominated by the imaginary part of $\sim$1 fm generated by
near-threshold resonances $f_{0}(980)$ and $a_{0}(980)$
\cite{ll1}.
It has been recently shown that the neglect of the FSI contribution
in the analysis of the two-$K_S^0$ correlation function
in Au+Au collisions at $\sqrt{s}_{NN}=200$ GeV would lead to
a noticeable ($\sim 25\%$) overestimation of the correlation
radius \cite{starksks}.

\section{FEMTOSCOPIC CORRELATIONS IN BETA-DECAY}

Let us now consider beta-decay of a nucleus with charge number
$Z_0$, four-momentum $p_0$, helicity $\lambda_0$
to a nucleus with charge number
$Z$, four-momentum $p$, helicity $\lambda$,
an electron (positron) with four-momentum
$p_{\rm e}$, helicity $\lambda_{\rm e}$ and
an antineutrino (neutrino) with four-momentum
$p_{\nu}$, helicity $\lambda_{\nu}$.
Taking into account the point-like character of the weak
interaction, the equal emission times of the decay particles and
the fact that the c.m. system of the electron and final nucleus
practically coincides with the rest frame of the initial nucleus
(i.e. ${\bf x}_{Z}\doteq {\bf x}_{Z_0}=0$ and
${\bf k}^*\doteq {\bf p}_{\rm e}$),
one can write the differential decay rate in the form:
\begin{eqnarray}
&d^5w \doteq \sum_{\lambda'{\rm s}}
\int d^3{\bf p}d^3{\bf p}_{\rm e}d^3{\bf p}_{\nu}
\delta^4(p_0-p-p_{\rm e}-p_{\nu})\cdot&
\nonumber\\
&
\cdot\left|\int d^3{\bf x}~{\cal T}({\bf x};
\lambda'{\rm s})
\exp(i{\bf p}_{\nu}{\bf x})
\psi_{-{\bf k}^*}^{(+)*}({\bf x}) \right|^2&
\nonumber\\
& \doteq \sum_{\lambda'{\rm s}}
\int d^3{\bf k}^*d^3{\bf p}_{\nu}
\delta(\omega_0-\omega-\omega_{\rm e}-\omega_{\nu})
\int d^3{\bf x}d^3{\bf x}'\cdot&
\nonumber\\
&\cdot {\cal T}({\bf x};\lambda'{\rm s})
{\cal T}^*({\bf x}';\lambda'{\rm s})
\exp[i{\bf p}_{\nu}({\bf x}-{\bf x}')]
\psi_{-{\bf k}^*}^{(+)*}({\bf x})\psi_{-{\bf k}^*}^{(+)}({\bf
x}'),\hskip.4cm&
\label{w}
\end{eqnarray}
where the amplitude ${\cal T}({\bf x};\lambda'{\rm s})$ is
basically determined by the distribution of the decaying
neutron (proton) within the nucleus.

It is instructive to consider the hypothetical situation when
the energy release in the decay is large and additional particles
are emitted. Then one could neglect the energy-momentum
conservation and get, after the integration over the neutrino
three-momentum (leading to delta-function
$\delta^3({\bf x}-{\bf x}')$), similar result
as in the case of multiparticle production on
conditions (i)-(iv):
\begin{equation}
\label{dw_hyp}
\frac{d^3w}{d^3{\bf k}^*} \propto
\int d^3{\bf x}\sum_{\lambda'{\rm s}}
\left|{\cal T}({\bf x};\lambda'{\rm s})
\psi_{-{\bf k}^*}^{(+)*}({\bf x})\right|^2
\propto \left\langle\left|
\psi_{-{\bf k}^*}^{(+)}({\bf x})\right|^2\right\rangle .
\end{equation}

The actual energy release in beta-decay is however
very small so that the integration over the neutrino
three-momentum does not lead to the diagonalization
of the spatial density matrix
$\sum_{\lambda'{\rm s}}{\cal T}({\bf x};\lambda'{\rm s})
{\cal T}^*({\bf x}';\lambda'{\rm s})$.
In fact, in so-called allowed
decays, the neutrino plane wave
$\exp(i{\bf p}_{\nu}{\bf x})$ can be even
substituted by unity.
Nevertheless, similar result as in Eq. (\ref{dw_hyp})
has been obtained by Fermi \cite{fer34} due to the fact
that the relativistic Coulomb wave function
$\psi_{-{\bf k}^*}^{(+)*}({\bf x})$ of the
electron (positron)-nucleus system changes very little
within the nucleus and can be taken out of the
integral in a form of the so-called Fermi function
${F}(k^*,Z,R)$. Neglecting the neutrino mass and the
nucleus recoil energy
(i.e. putting $\omega_\nu=|{\bf p}_\nu|$ and $\omega=M$),
one can write in the nucleus rest frame ($\omega_0=M_0$):
\begin{equation}
\label{dw}
\frac{d^3w}{d^3{\bf k}^*} \doteq 4\pi {F}(k^*,Z,R)
(M_0-M-\omega_e)^2
\sum_{\lambda'{\rm s}}
\left|\int d^3{\bf x}~
{\cal T}({\bf x};\lambda'{\rm s})\right|^2 ,
\end{equation}
where
\begin{eqnarray}
&&{F}(k^*,Z,R)\doteq
\langle |\psi_{-{\bf k}^{*}}^{(+)}({\bf r}^{*})|^{2}
\rangle \doteq |\psi_{-{\bf k}^{*}}^{(+)}(R)|^{2}
\nonumber\\
&&\hskip.3cm\doteq (2k^*R)^{2\sigma}\frac{2\sigma+4}
{[\Gamma(2\sigma+3)]^2}
\exp(-\pi\eta)|\Gamma(\sigma+1-i\eta)|^2 ,~~~~~~
\label{FF}
\end{eqnarray}
$\eta\doteq \mp Ze^2\omega_e/k^*$
for electrons (positrons), $\sigma=(1-Ze^2)^{1/2}-1$.
The substitution of the separation $r^*$ by the nucleus radius $R$
in the second equality in Eq. (\ref{FF})
is justified due to a weak $r^*$-dependence of the wave function
within the nucleus \cite{fer34}; the third equality neglects the
screening effect of the atomic electrons.

One may see that the nucleus radius enters in the
Fermi function through the factor $(2k^*R)^{2\sigma}$ which is essentially
different from unity only for sufficiently large charge numbers $Z$.
At small $Z$-values, $\sigma\doteq 0$ and the Fermi function reduces
to the Coulomb penetration (Gamow) factor
$A_c(\eta)=|\psi_{-{\bf k}^{*}}^{(+)}(0)|^{2}=
2\pi\eta[\exp(2\pi\eta)-1]^{-1}$.
The sensitivity of the Fermi function
to the nucleus radius is demonstrated in figure 3.

\begin{figure}[!htb1]
\begin{center}
\includegraphics*[angle=0, width=8cm]{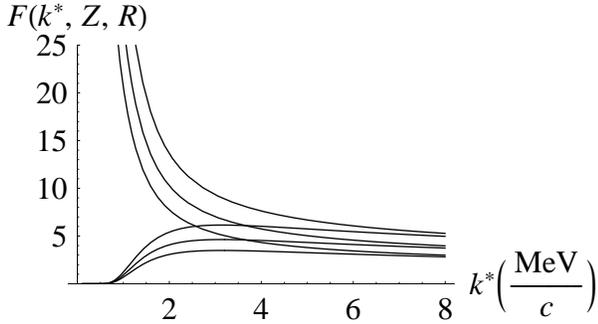}
\end{center}
\caption{\emph{ \small The Fermi function $F(k^*,Z,R)$
for beta-decay to a final nucleus of charge $Z=83$
as a function of the electron (decreasing curves) or positron
momentum $k^*$ and the
nucleus radius $R=2,4,8$ {\rm fm} (in decreasing order).}}
\end{figure}

\section{Conclusions}

We have considered femtoscopic QS and FSI momentum
correlations in multiparticle production and beta-decay,
as well as spectroscopic QS space-time correlations in the
detected radiation from a distant source.
We have demonstrated the orthogonal character of the
femtoscopic and spectroscopic correlations,
earlier pointed out by Kopylov and Podgoretsky
\cite{KP75}.
We have shown that the same
functional form of the two-particle
correlation function in multiparticle production and
Fermi function in beta-decay (both being equal to the
average square of the two-particle wave function)
is due to different reasons.
In former case, this result
is valid in the approximation of independent classical
quasi-point-like particle emitters and sufficiently
small freeze-out phase space density, while in the latter
case it follows from a weak variation of the
electron (positron)-nucleus wave
function within the nucleus volume and a point-like
character of beta-decay.
It should be stressed that a small space-time extent of the
emitters alone does not guarantee the validity of the
approximation of classical emitters (the diagonalization
of the space-time density matrix).
This approximation
may naturally be justified in high-energy multiparticle processes
due to the minor importance of conservation laws in this case.

\bigskip
\noindent\textbf{Acknowledgements}
This work was supported by the
Grant Agency of the Czech Republic under contract
202/07/0079
and partly carried out within the scope of the GDRE:
Heavy ions at ultrarelativistic energies -–
a European Research Group comprising
IN2P3/CNRS, EMN, University of Nantes,
Warsaw University of Technology, JINR Dubna, ITEP Moscow and
BITP Kiev.

\end{document}